\documentclass{iopconfser}
\UseRawInputEncoding
\usepackage{bbm}
\usepackage[utf8]{inputenc}
\usepackage{placeins}
\usepackage{amssymb,amsmath,amsfonts}
\usepackage[colorlinks,linkcolor=purple,citecolor=teal]{hyperref}
\usepackage{dsfont}
\usepackage{color}
\usepackage{graphicx}
\usepackage{hyphenat}
\usepackage{wrapfig}
\usepackage{empheq}
\usepackage{textcomp}
\usepackage[caption=false]{subfig}
\usepackage{rotating}
\usepackage{wrapfig}
\usepackage{pdfpages}
\usepackage{verbatim}
\usepackage{setspace}
\usepackage{array}
\usepackage{caption}
\usepackage[pdf]{pstricks}
\usepackage{upgreek}
\usepackage{cmll}
\usepackage{latexsym}
\usepackage{braket}
\usepackage{cite}
\usepackage{epsfig}
\usepackage[left=2.4cm,top=3.3cm,right=2.4cm,bottom=3.3cm,bindingoffset=0cm]{geometry}
\usepackage[titletoc,page]{appendix}
\usepackage{multicol}
\usepackage{youngtab}

\newcommand{\beq}{\begin{eqnarray}}
\newcommand{\eeq}{\end{eqnarray}}
\newcommand{\bea}{\begin{eqnarray}}
\newcommand{\eea}{\end{eqnarray}}
\newcommand{\be}{\begin{equation}}
\newcommand{\ee}{\end{equation}}

\def\brc{\langle}
\def\ckt{\rangle}

\def\de{\partial}

\numberwithin{equation}{section}

\numberwithin{equation}{section}

\begin{document}

\title{The Quantum Ratio\footnote{An invited talk by K.K. presented at  DICE 2024, Castiglioncello, Italy.  }}

\author{K Konishi$^{1,2}$,   
H - T  Elze$^{1}$}

\affil{$^1$Department of Physics, University of Pisa, Pisa, Italy}
\affil{$^2$ INFN, Sezione di Pisa, Pisa, Italy}

\email{kenichi.konishi@unipi.it, elze@df.unipi.it}

\begin{abstract}
The concept of the Quantum Ratio was born out of the efforts to find a simple but universal criterion if the center of mass (CM) of  an isolated (microscopic or macroscopic) body behaves quantum mechanically or classically, and under which conditions. It  is defined as the ratio 
between the quantum fluctuation range, which  is the spatial extension of the pure-state CM  wave function, and the linear size of the body (the space support of the internal, bound-state wave function).  The two cases where the ratio is smaller than unity or 
  much larger than unity, roughly  correspond to  
 the body's CM behaving classically or quantum mechanically, respectively. 
  An important notion following from the introduction of quantum ratio is that the elementary particles  (thus the electron and the photon) are quantum mechanical.  This is so  even when the environment-induced decoherence turns them into a mixed state. Decoherence  (mixed state) and classical state should not be identified. This simple observation is further elaborated, by analyzing some atomic or molecular processes.   It 
   may have far-reaching implications on the way quantum mechanics works, e.g.,  in biological systems.

\end{abstract}

\section{The Quantum Ratio}

\begin{figure}
\begin{center}
\includegraphics[width=3in]{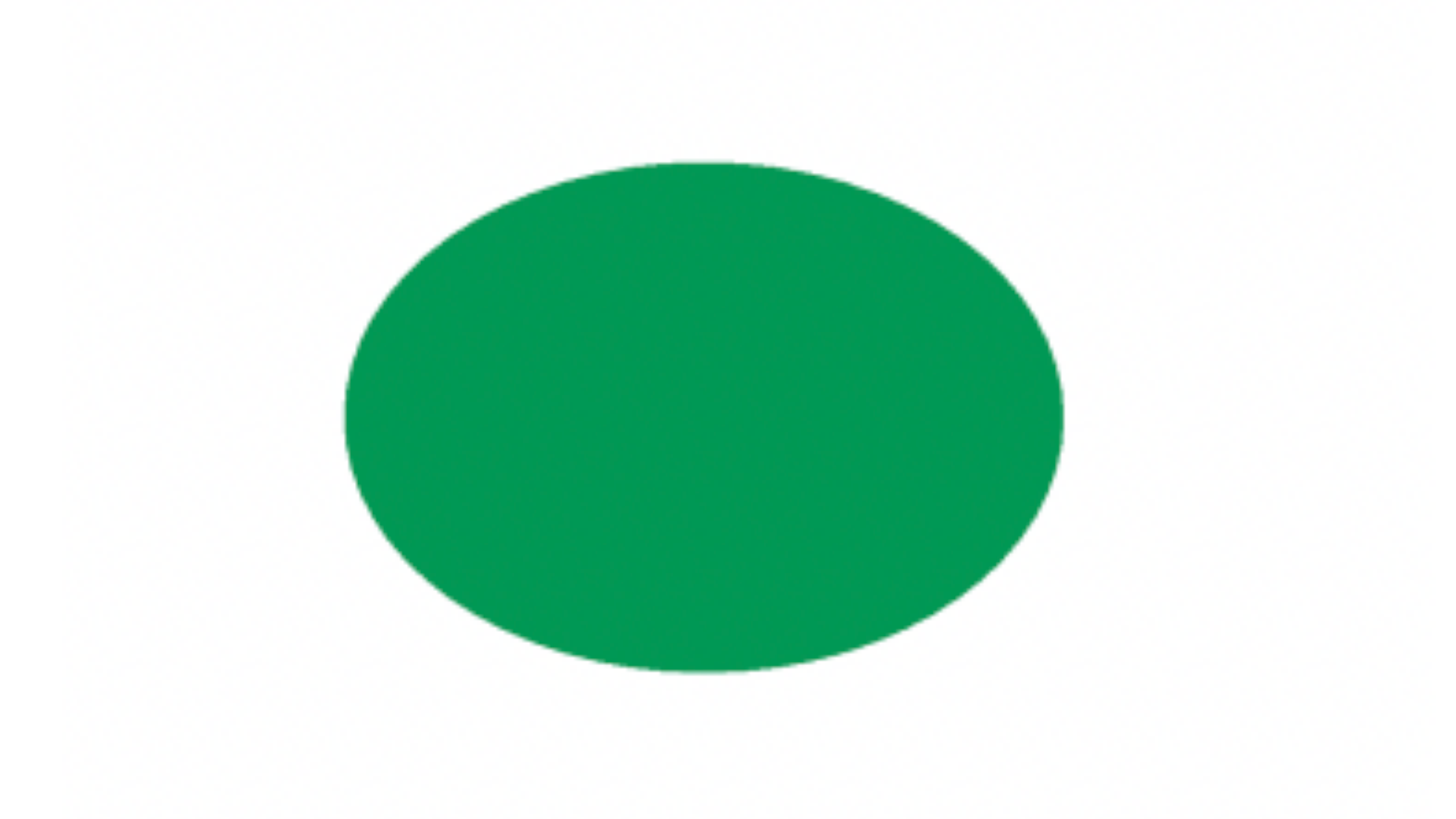}
\caption{Does this body behave classically or quantum mechanically? }
\label{Question}
\end{center}
\end{figure}

The question we are going to discuss in this talk is this: given an isolated microscopic, mesoscopic or macroscopic body,  see Fig.~\ref{Question},   does its center of mass (CM) behave classically or quantum mechanically? 
The question might sound deceptively simple, or perhaps, poorly defined. Actually, an attempt to answer it will eventually take us  to the entire issues of what might be called the Great Twin Puzzles of Physics Today, namely,   (i)   the so-called Quantum Measurement Problems on the one hand, and (ii)  how Newton's law for macroscopic bodies emerges from quantum mechanics, on the other.  The first of them  was addressed recently in \cite{KK1,KKTalk,KK3}.
The second problem was investigated in   \cite{KK2,KKHTE,KM}.

The concept of Quantum Ratio summarizes, in an approximate but universal way, how to discriminate whether an isolated body is best described by quantum mechanics or by Newton's equations. It is defined by
\be    Q \equiv   \frac{R_q}{L_0}\;,    \label{QR}
\ee
where $R_q$ is the quantum fluctuation range  of the CM of the body,  and $L_0$ is  its  (linear) size.  
The criterion proposed to tell whether the body behaves quantum mechanically or  classically   is     \cite{KK2,KKHTE} 
\be  Q    \gg 1\;, \qquad  {\rm (quantum)},
 \label{quantum}   \ee
 or 
  \be      Q   \lesssim   1    \;,\qquad  {\rm (classical)}\;,     \label{classical} 
 \ee
 respectively.

 Let us take the total wave function of the body in a factorized form, 
\be   \Psi({\bf r}_1,  {\bf r}_2, \ldots  {\bf r}_N) =   \Psi_{CM}({\bf R}) \,\psi_{int}({\hat  {\bf r}}_1,  {\hat  {\bf r}}_2, \ldots   {\hat  {\bf r}}_{N-1}) \;.   \label{factorization}
\ee
where $  \Psi_{CM}$ is the CM wave function,  the $N$-body bound state is described by the internal wave function $\psi_{int}$. $\{ {\hat  {\bf r}}_1,  {\hat  {\bf r}}_2, \ldots   {\hat  {\bf r}}_{N-1}\}$  are the internal positions of the component atoms or molecules, ${\bf R}$ is the CM position and    ${\bf r}_i =  {\bf R}+  {\hat  {\bf r}}_i$  ($i=1,2,\ldots,N$).      In the case of a macroscopic body $N$ can be as large as $N\sim 10^{25}, 10^{50},$  etc.  The size of the body can be defined as 
\be    L_0  =  {\rm Max}_i \,  {\bar  r_i}\;, \qquad       {\bar  r_i} \equiv (\brc  \psi_{int} | ({\hat {\bf r}}_i)^2   | \psi_{int} \ckt )^{1/2} \;, \label{sizeL}
\ee
whereas the quantum range $R_q$ is simply the spatial extension of the pure-state CM wave function $\Psi_{CM}$.
A few remarks:
\begin{description}
  \item[(i)]   There are no a priori upper limit on  $R_q$:  this is closely related to the well-known quantum nonlocality.  This originates from the fact that QM has 
  no fundamental constant with the dimension of a length \cite{KK1}.  Note also that even the normalization condition of a single particle wave function, $||\psi||=1$,   does not limit  
   $R_q$ in general  (recall Weyl's criterion). 
  
  \item[(ii)]   $R_q$ is restricted by decoherence,  it depends on the body temperature (for an isolated macroscopic body), or on the environment.
  
  \item[(iii)]   The size of the wave packet of the CM wave function, $\Delta_{CM}$,  should not be identified either  with  $L_0$ or  with   $R_q$.  Being a measure of a spread of the wave function, however, it does mean
 \be    R_q \gtrsim  \Delta_{CM} \;,     \label{together}     \ee        
but  $R_q$ can be much larger than  $\Delta_{CM}$.  
  
  \item[(iv)]  $L_0$ and $\Delta_{CM}$ are independent of each other:  given a body with size $L_0$, its CM can have a wave function with a narrow wavepacket,
 $\Delta_{CM}\ll  L_0$,  or  with spread  much larger than it:    $\Delta_{CM} \gg  L_0$.  
 
\end{description}

Also, the wave packet of a free particle   diffuses in time. The diffusion time (which may be defined appropriately) depends on mass in an essential way, see Table~\ref{diffusion}. 
 \begin{table}
  \centering 
  \begin{tabular}{|c|c|c|  }
\hline
 particle   &   mass  (in $g$)  &    diffusion time   (in $s$)  \\   \hline
 electron   &   $9  \cdot  10 ^{-28}  $  &     $10^{-8}   $   \\
   hydrogen atom   &   $1.6  \cdot  10 ^{-24}  $    &    $1.6 \cdot 10^{-5}  $  \\
   $C_{70}$ fullerene  &   $8   \cdot  10 ^{-22} $  &     $8 \cdot 10^{-3}$    \\
   a stone of $1g$    &     $  1  $    &     $10^{19}      $  \\  
\hline
\end{tabular}
  \caption{ \footnotesize  Diffusion  time  of the free wave packet for different particles. Here we take the initial wave packet size of $\Delta_{CM} = 1 \mu =10^{-6} m$, and define 
  the diffusion time  as $\Delta t$ needed for doubling it.  For a  macroscopic particle of $1 g$, the doubling time,  $   10^{19}  {\rm sec}   \sim 10^{11}  {\rm yrs}$,
   exceeds  the age of the universe.
       }\label{diffusion}
\end{table}

\subsection{Warning} 

Even if the CM of a macroscopic body might behave classically, the microscopic degrees of freedom inside the body are always quantum mechanical
(see the discussion below),  
a fact fundamental in 
biological processes \cite{QuantumBiology}.

\section{Quantum Ratio: illustration}

\subsection{Elementary particles}

For elementary particles, 
\be   L_0=0\;,  \qquad    .^.. \qquad     Q=\infty\;. 
\ee
The elementary particles are quantum mechanical.  The elementary particles known today are the quarks, leptons, the gauge bosons (the photon, $W$ and $Z$
bosons, and the Higgs scalar (Appendix~\ref{EP}).   Note that the fact that the world is described very precisely by the so-called standard quantum field theory of these elementary particles, 
$SU(3)_{QCD} \times (SU(2)\times U(1))_{GWS}$ (known as the Quantum ChromoDynamics and Glashow-Weinberg-Salam electroweak theory) \cite{Weinberg,Salam,Glashow,GellMann}, up to the energies $O(10)\,$TeV,     means  that
    \be   L_0   \lesssim  O(10^{-18})\, {\rm cm}\;.    \label{elempart}    \ee      
    It can be taken to be $0$ for any physics purpose at the nuclear, atomic or larger distance scales.  
    
 A familiar idea  in physics is that the size of an object is a relative concept.  Any object may look pointlike,  if 
 observed  from a much greater distance than its size.  Indeed, this concept survives in a subtle and precise way  in  {\it  renormalizable quantum field theories}  such as the the standard $SU(3)_{QCD} \times (SU(2)\times U(1))_{GWS}$ theory. Namely the system is invariant under renormalization group (RG) \cite{Wilson2}:  that is,  physics looks alike when the relevant scale is changed, as long as the coupling constants are appropriately varied (i.e.,  obeying the  RG equations). In a sense, therefore, the system is invariant under dilatations  \cite{Coleman}. 

However, this scale invariance  is broken by the vacuum expectation value of the Higgs scalar, $ \brc  \phi^0 \ckt \simeq   246 \,   {\rm GeV}$,   and by the RG invariant mass scale of QCD \footnote{This is the mass scale at which the coupling constant of QCD becomes strong.},  $\Lambda_{QCD}\simeq   250 \,   {\rm MeV}$.   All the mass parameters (Appendix~\ref{EP}) of our world \cite{PDG}  arise from the above two  and from some dimensionless coupling constants in the $SU(3)_{QCD} \times (SU(2)\times U(1))_{GWS}$
theory. 

In other words,  the world we live in have definite characteristic scales (such as the Bohr radius, and the size of the nuclei). Accordingly, the concepts such as the microscopic (nuclear, atomic, molecular) or   macroscopic  (much larger than those) systems,  have a well-defined, concrete  meaning. 

\subsection{Atomic nuclei and hadrons }

The atomic nuclei and the hadrons  ($p$, $n$, $\pi^{\pm}$, $\pi^0$, etc.)  have all sizes of the order of  $L_0 \sim 1$ fm,  that is $O(10^{-13})$ cm.

\subsection{Atoms} 

The atoms have a characteristic size of the order of 
\be   L_0   =   0.5  \sim  10^2  \,    \AA\;.
\ee
In the famous Stern-Gerlach experiment \cite{SG},  a silver atom of size  $L_0 \sim 1.4 \, \AA$ is sent into a region of magnetic field of strong gradient.
Its wave packet (of size  $0.02-0.03$  mm)  splits into two subpackets, separated by distances $\sim 0.2$ mm.    The spatial support of the wave function $\psi$ can be taken to be about this size.    It follows that 
\be     Q =    \frac{R_q}{L_0} \gtrsim    \frac{0.2 \, mm}{1.4 \, \AA} \simeq  10^6  \gg 1\,.
\ee
The silver atom,  a quantum-mechanical bound state of $47$ electrons, $47$ protons and $51$ neutrons and 
with mass $\sim 100$ times that of the hydrogen atom,  thus behaves perfectly as a quantum mechanical particle,  as a whole.

\subsection{Molecular interferometry}   

\begin{figure}
\begin{center}
\includegraphics[width=4in]{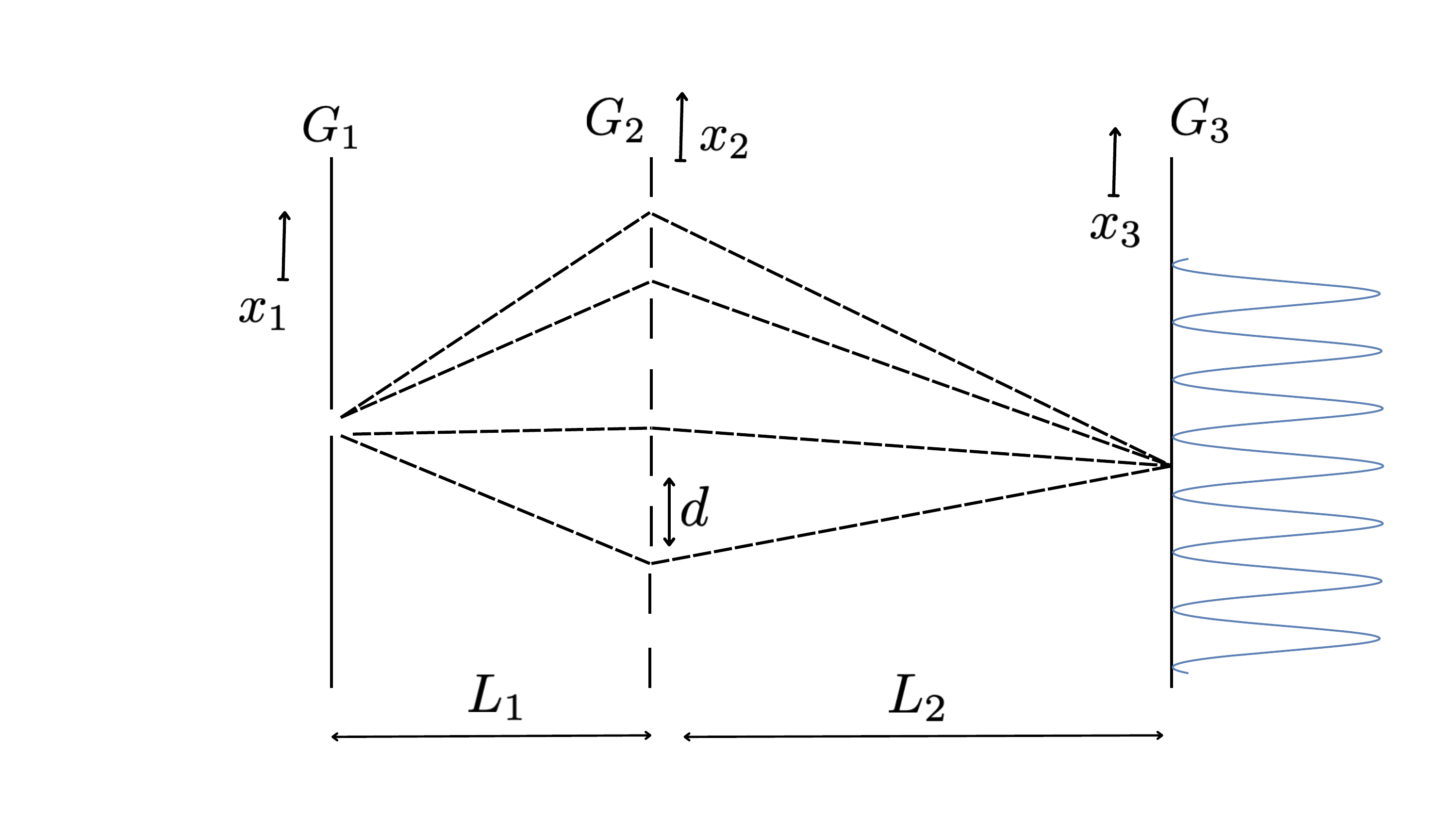}
\caption{\footnotesize The Talbot effect \protect{\cite{Talbot}}.  The intensity modulation of the molecules  immediately after the passage of the diffraction grating $G_2$, Fig.~\ref{modulation}, is reproduced,  due to the sum over paths,  at an imaging plane $G_3$ placed at definite distances  $L_2$, related to the Talbot length  
$ L_{T}  =    \frac{d^2}{\lambda_{dB}}\; $ from $G_2$.   }
\label{Talbotimage}  
\end{center}
\end{figure}

\begin{figure}
\begin{center}
\includegraphics[width=3.5in]{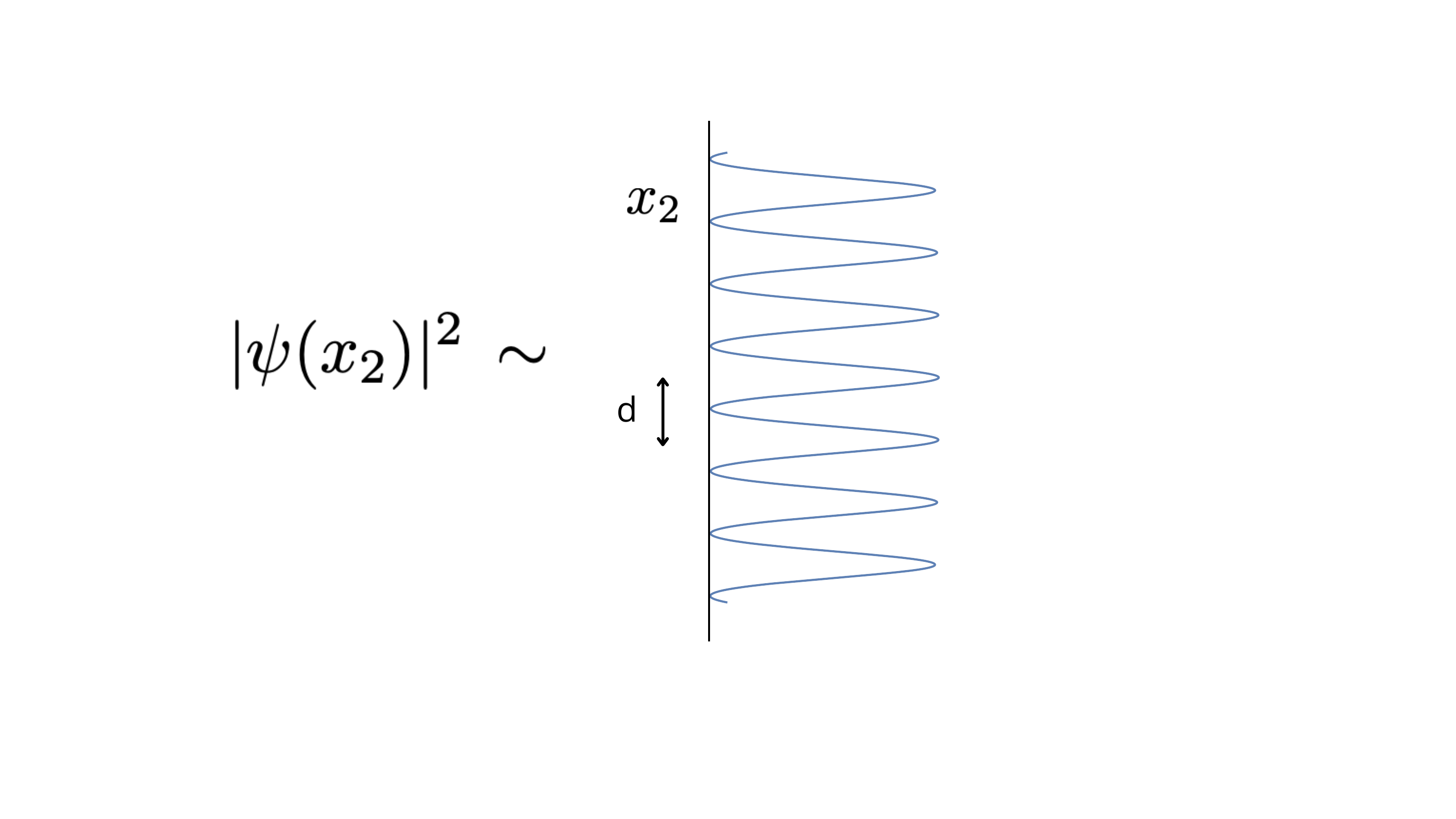}
\caption{\footnotesize   The intensity pattern of the atom (or molecule) behind the diffraction grating  $G_2$.  Each peak corresponds to a slit opening.  }
\label{modulation}
\end{center}
\end{figure}

\begin{figure}
\begin{center}
\includegraphics[width=4.0in]{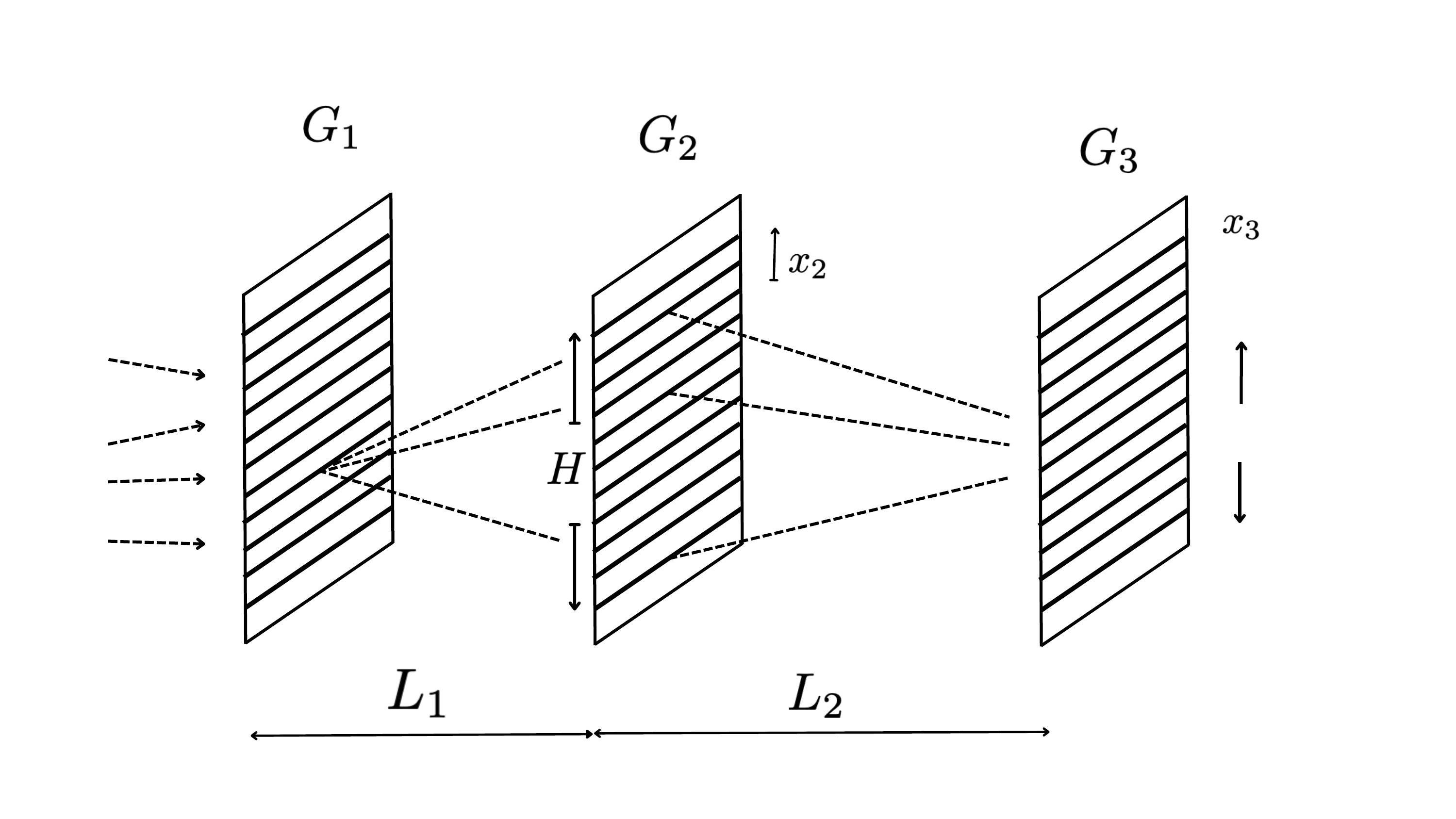}
\caption{\footnotesize   Talbot-Lau interferometer.  $G_2$ is the diffraction grating. Thick lines are the slit openings.  $G_3$ is a transmission-scanning grating movable vertically.   $G_1$ are  the  source slits.  }
\label{TalbotLau}
\end{center}
\end{figure}

Many beautiful atomic or  molecular interference experiments have been performed in recent years \cite{Keith}-\cite{Bateman}.  Many of them makes use of the Talbot-Lau interferometry, 
illustrated schematically  in Fig.~\ref{Talbotimage},  Fig.~\ref{modulation},  Fig.~\ref{TalbotLau}.   

The rough estimate of the quantum ratio for the atom or molecule in these experiments is discussed in  \cite{KKHTE}: the result is shown in Table~\ref{QuantumRatio}.  

\begin{table}  \centering 
  \begin{tabular}{|c|c|c|c|c|c|c|}
\hline
Particle &  mass    &    $L_0$  &    $R_q$   &    $Q $ &   Exp  &   Miscl  \\
\hline
  $Ag$   &  $108 $     &   $1.44 $   &    $0.2$  &    $\sim 10^6$  &      \cite{SG}   &   Stern-Gerlach   \\  
  $Na$   &   $23 $  &  $2.27$  &    $0. 5 / 0.75 $  &   $\sim 10^6$  &      \cite{Chapman}     &  \\  
      $C_{70}$   &  $840$    &      $9.4$       &    $16$    &    $\sim  10^7$  &     \cite{C70Bis,BrezArndtZeil,C70}  &       $T \ll 2000K$       \\     
            $C_{70}$   &  $840$    &        $9.4$      &    $\sim  0.001$        &    $ \sim  10^3  $        &    \cite{C70}   &     $T \ge 3000K$   \\                
\hline
\end{tabular}
  \caption{\footnotesize  The  size ($L_0$), the quantum fluctuation range ($R_q$)   and the 
  quantum ratio     $Q \equiv  R_q/L_0$   of atoms and molecules   in various  experiments.  The mass is in atomic unit  ($au$);  $L_0$ is given in Angstr\"om  ($\AA$);   $R_q$  is in $mm$.  In all cases,  the momentum of the atom (molecule), their masses, the size of the whole experimental apparatus, thus the time interval involved, 
are such that the  quantum diffusion of their (transverse) wave packets are negligible.
      }  \label{QuantumRatio} 
\end{table}

\subsubsection{Remark on ``matter wave''} 

A familiar expression used often in the articles on the atomic and molecular interferometry  \cite{Brand}-\cite{Bateman}  is  ``matter wave".  
It might appear to summarize nicely  the characteristic feature of quantum-mechanics:  ``wave-particle duality". 
Actually,   such an expression is more likely to obscure the essential quantum mechanical features of these processes, rather than illuminating  them. 
 It appears to imply that the beams of atoms or molecules  somehow behave as a sort of wave:  this is not an accurate description of the processes studied.   The wave-particle duality of de Broglie, the core concept of quantum mechanics, is  the property of  {\it  each single quantum-mechanical particle,}  and not of any unspecified  collective motion of particles in the beam  \footnote{The ``wave nature" of atoms or molecules 
 observed in  the  interferometry  \cite{Brand}-\cite{Bateman}  must   be distinguished from  the  many-body collective quantum phenomena, such as Bose-Einstein condensed ultra cold atoms described by a macroscopic wave function.}.
  This point was demonstrated experimentally by Tonomura et. al.\cite{Tonomura} in a double-slit electron  interferometry experiment   \`a la Young,  with exemplary clarity.   

 Exactly the same phenomena occur in any atomic or molecular interferometry.  
As the correlation among the  atoms or molecules  in the beam is  negligible   (as  it should be),  and the position of each final atom/molecule is apparently random,  the resulting interference fringes  such as manifested in  the Talbot (or the Talbot-Lau) interferometers, is all the more surprising and interesting.  What these experiments  show  goes much deeper  into the heart of QM,   
 than  the  words,  ``matter wave" or ``wave-particle duality",  might suggest.

\subsection{A reflection}

Thus the electron, being an elementary particle, is quantum mechanical ($L_0=0,$ $Q=\infty$). 
 On the other hand, it is well known that an electron decoheres in $10^{-13}$ sec,    
in the $300K$  atmosphere at $1$ atm pressure \cite{Joos1}-\cite{Zurek2}.   So what is happening?   The only sensible conclusion to draw is that 
decoherence and classical limit are two distinct concepts: they should not be identified.  Decoherence does not mean in itself  that the particle affected becomes classical, even though the classical behavior of macroscopic bodies do require decoherence (see Appendix~\ref{NewtonEq}).

\section{Decoherence does not imply classical}

This observation, which follows at once from the concept of  Quantum Ratio applied  to the elementary particles,  can have far-reaching consequences.  Let us discuss this question with a few atomic or molecular processes.

\subsection{Stern-Gerlach processes with small and large spins}

First  we discuss the SG  process again,  in more detail,  in three different regimes,  (i)  a pure QM process;  (ii)  the environmental decoherence  (an incoherent, mixed state);  and (iii)  for a classical particle. The main aim is to highlight  the differences between these different physics situations as sharply as possible.   

\subsubsection{Pure  spin $\frac{1}{2}$  state  \label{pure} }

In the Stern-Gerlach experiment for a spin $\frac{1}{2}$ particle, the  wave function 
\be    
  \Psi =  \psi_1({\bf r}, t) |\!\uparrow\ckt +   \, \psi_2({\bf r}, t)   |\!\downarrow\ckt \;.   \label{wavepackets}
\ee
splits, in an inhomogeneous magnetic field, into two subpackets,  each obeying the Schr\"odinger equation, 
\be   i \hbar  \frac{\de}{\de t}   \psi_{1,2}  =     \left(    \frac{{\bf p}^2}{2m}   \pm   \mu_{B}   B_z  \right)       \psi_{1,2} \;.     \label{SEq1}
\ee
For certain subtleties in the  Stern-Gerlach processes, see Appendix~\ref{SGB} and  \cite{Alstrom,Platt}.

From (\ref{SEq1}) and their complex conjugates,  the Ehrenfest theorems for spin-up and spin-down components  follow  separately,  
      \bea  &&   \frac{d}{dt}   \brc {\bf r}\ckt_1   =  \brc {\bf p}/{m} \ckt_1  \;,   \quad   
      \frac{d}{dt}  \brc {\bf p}\ckt_1  =   -    \brc  \nabla (\mu_{B}   B(z)) \ckt_1\;;      \label{New1}   \\
   &&    \frac{d}{dt}   \brc {\bf r}\ckt_2   =  \brc {\bf p}/{m} \ckt_2  \;,   \quad   
      \frac{d}{dt}  \brc {\bf p}\ckt_2  =   +  \brc  \nabla (\mu_{B}   B(z)) \ckt_2  \;,   \label{New2}
\eea
where    $\brc {\bf r}\ckt_1  \equiv   \brc   \psi_1 | {\bf r} | \psi_1 \ckt$, etc.       Nevertheless, the two subwavepackets $\psi_1$ and $\psi_2$
remain in coherent superposition.

\begin{figure}
\begin{center}
\includegraphics[width=5in]{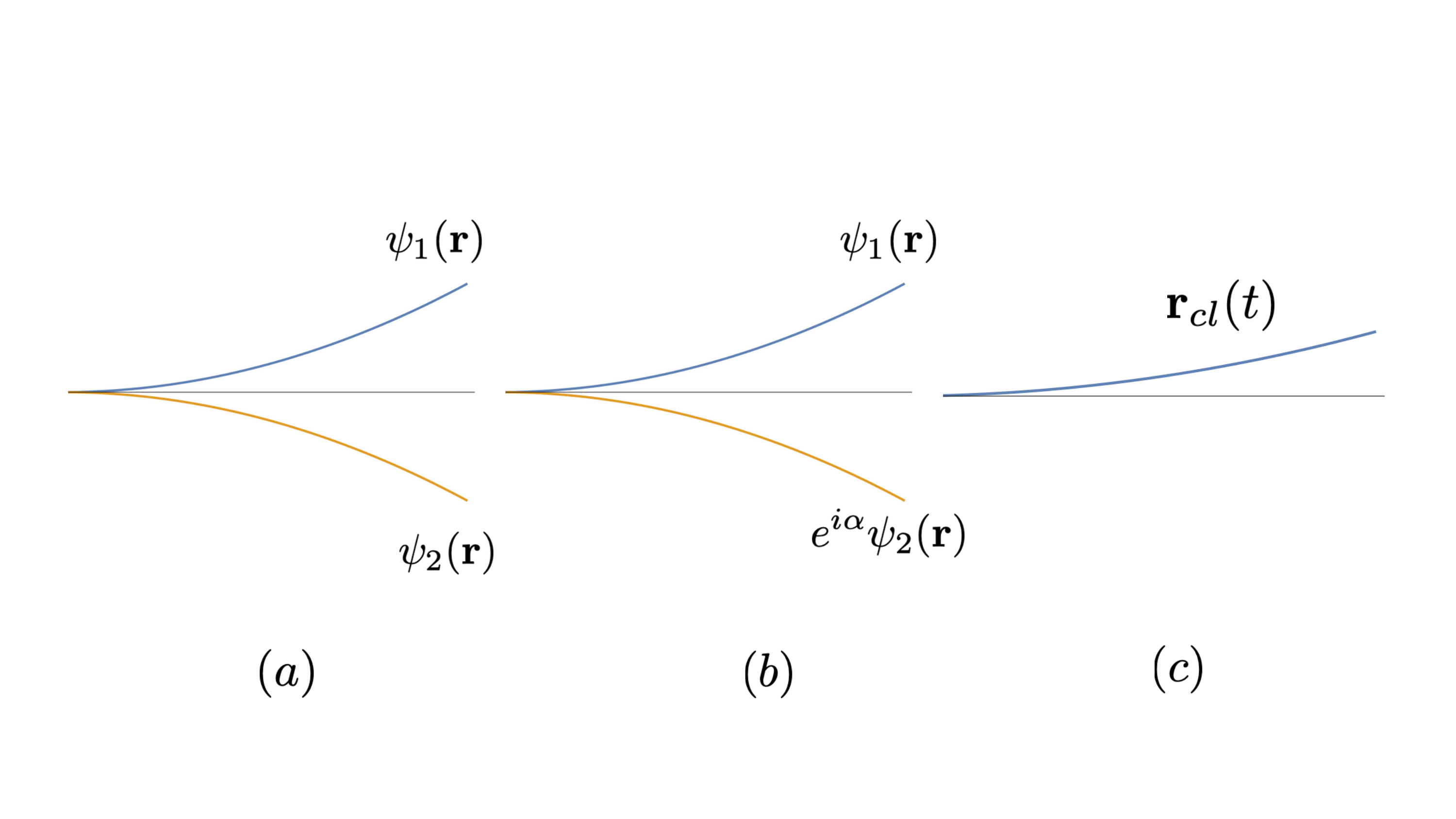}
\caption{\footnotesize  The spin-up and spin-down sub wavepackets of the silver atom evolve independently under the Schr\"odinger equation, both in the vacuum (fig.(a),  pure state)
and in a weak environment (\ref{cond1})-(\ref{cond3})   (fig.(b)),  where the decoherence is represented by an unknown, to-be-averaged-over,  relative phase $\alpha$ between $\psi_1({\mathbf r})$ and $\psi_2({\mathbf r})$.  Fig.(c)  represents a unique classical trajectory.     }
\label{WSG}
\end{center}
\end{figure}

\subsubsection{Spin $\frac{1}{2}$   and weak decoherence  \label{decoh1} }

When the system is immersed in an environment,  it rapidly decoheres.  The density matrix in the position representation gets reduced 
  at times   $t\gg 1/\Lambda$, to   a  diagonal  form  
\be    \psi({\bf r})  \psi({\bf r}^{\prime})^*  \to    \psi_1({\bf r})     \psi_1({\bf r}^{\prime})^*  |\!\uparrow\ckt \brc \uparrow|    +   \psi_2({\bf r})     \psi_2({\bf r}^{\prime})^* 
    |\!\downarrow\ckt \brc \downarrow| \;,\qquad   |{\bf r}_1-{\bf  r}_2| \gg  \lambda\;,     \label{mixed} 
  \ee
  where $\Lambda$ is the decoherence rate  \cite{Joos1}-\cite{Zurek2} and $\lambda$ is the de Broglie wavelength of the environment particles.  The diagonal density matrix  (\ref{mixed}) means that each atom is now   {\it    either near ${\bf r}_1$ or  near   ${\bf r}_2$.}
   The prediction for the SG experiment is however similar to  the case of spin-mixed state:   it cannot be distinguished from the prediction $|c_1|^2 : |c_2|^2$ for the relative intensities of the two image bands in the case of the pure state. 
  
 Actually, the study of the effects of the environment particles  is  a complex, and highly nontrivial problem,   as it involves many factors such as the density and flux of these particles,  the pressure, the average temperature,  kinds of the particles present and the type of interactions, and so on   \cite{Joos1}-\cite{Zurek2}.    A simple statement such as (\ref{mixed}) might sound as  an oversimplification.

Without going into details, we may nevertheless enlist  the basic conditions under which the result  (\ref{mixed}) can be considered reliable.  
Following  \cite{Tegmark},  we  introduce  the {\it  decoherence time}  $\tau_{dec} \sim 1/ \Lambda_{dec}$,   as a typical timescale over which the decoherence takes place.  Also the {\it  dissipation time}  $\tau_{diss}$
may be considered, as a  timescale in which the loss of the energy, momentum  of the atom under study due to the interactions with the environmental particles, become significant  \footnote{
Unlike \cite{Tegmark}, however, we do not consider $\tau_{dyn}$, the typical timescale of the internal motion of the object under study. 
Roughly speaking the size $L_0$ (the space support of the internal wave function) we introduced in defining  the quantum ratio,  (\ref{QR}),   corresponds to it ($\tau_{dyn}   \propto  L_0$).
Quantum-classical criteria   suggested by \cite{Tegmark} might appear to have some similarity with (\ref{quantum}), (\ref{classical}).  However,  the former seems to leave unanswered questions such as ``what happens to a quantum particle ($\tau_{dyn} < \tau_{dec}$), at  $t> \tau_{dec}$?"     This is precisely the  sort of question we are trying to address here. 
}.  
We need to consider also a typical {\it  quantum diffusion time},   $\tau_{diff}$, and finally, the {\it  transition time},    $\tau_{trans}$,  the interval of time the atom  spends between the source slit to the 
image screen.   
Summarizing,  we consider the time scales 
\be    \tau_{dec} \ll  \tau_{trans}  \ll    \tau_{diff}, \tau_{diss}\;. \label{cond1}
\ee   
The first inequality tells that the motion of the wave packets is much slower than  the typical decoherence time.  Consider the atom at some point, where  it is described
by a split wave packet of the form (\ref{wavepackets}),  with their centers separated   by 
\be   |{\bf r}_1 - {\bf r}_2|  \gg  a\;,      \label{cond2}   \ee
   where $a$ is the size of the original wavepacket.  We may then treat such an atom as if it were  
 at rest,  and take into account the rapid decoherence processes studied in \cite{Joos1}-\cite{Zurek2} first  (a sort of Born-Oppenheimer approximation).
Furthermore,  let us  also take the typical de Broglie wavelength  $\lambda$  of the environment particles  such that 
\be     a \ll     \lambda    \ll       |{\bf r}_1 - {\bf r}_2| \;.   \label{cond3}
\ee
Namely,  the environment particles can resolve between the split wave packets,  but not the interior of  each of the subpackets,    $\psi_1({\bf r})$ or     $\psi_2({\bf r}).$

Under the conditions  (\ref{cond1})-(\ref{cond3}),  each of the split wave packets proceeds just  as in the   pure case (no environment) reviewed in Sec.~\ref{pure},   whose average position and momentum  (i.e., the expectation values)  obey Newton's equations, 
  (\ref{New1}), (\ref{New2}).   
  Each of the subpackets describes a quantum particle, in a (position) mixed state,  that is, either near  ${\bf r}_1$
or  ${\bf r}_2$.  After leaving the region of the SG magnets, it is just a (pure-state) wave packet  $\psi_1({\bf r})$ or   $\psi_2({\bf r})$.  The  two wave functions  however  no longer interfere, Fig.~\ref{WSG} (b),    in contrast to the pure split wave packet studied in Sec.~\ref{pure}, 
Fig.~\ref{WSG} (a).   

Note that  if any of the conditions (\ref{cond1})-(\ref{cond3}) are violated  the motion of the atom would be very different.  For instance,  $\tau_{diss} \ll  \tau_{trans}$  would mean a totally random motion for the atom. Even in such a case,  though, the effects of the environment-induced decoherence/disturbance are quite distinct from that of a classical motion of a particle, with a unique, well-defined  trajectory, 
discussed below,   Fig.~\ref{WSG} (c).

  \subsubsection{Classical  particle} 
  
A classical particle, with  the magnetic moment directed   towards    
\be  {\bf n}=  (\sin \theta  \cos \phi,  \sin \theta  \sin \phi,   \cos \theta)\;, \ee
   is described by  Newton's  equation, 
\be   m\, {\dot {\bf r}} = {\bf p}\;, \qquad  \frac{d p_x}{dt} = \frac{d p_y}{dt} = 0\;, \quad         \frac{d p_z}{dt} =    F_z  =  - \frac{\partial}{\partial z}   {\boldsymbol {\mu}  \cdot {\bf B}}\;. \label{Newton} 
\ee
It traces a unique, well defined trajectory   (Fig.~\ref{WSG} c). 
The way  the unique trajectory for a classical particle emerges from quantum mechanics has been discussed in \cite{KK2},  where the magnetic moment is an expectation value 
\be          \sum_i   \langle \Psi |    ( {\hat   {\boldsymbol \mu}}_i +     \frac{e_i    {\hat  {\boldsymbol \ell}}_i  }{2  m_i  c } ) |\Psi \rangle  =   {\boldsymbol  \mu}\;,    \label{classicalSG}
\ee
taken in the internal bound-state wave function $\Psi$   and $\mu_i$ and    $\frac{e_i    {\bf \ell}_i  }{2  m_i  c }$ denote the intrinsic magnetic moment and one due to the orbital motion of the $i$-th constituent atom (molecule);    $i=1,2,\ldots, N$.  Clearly,   in general,  the considerations made in  Sec.~\ref{pure} and  Sec.~\ref{decoh1} for a spin $1/2$ atom,  with a doubly split wave packet,    cannot be  generalized   simply to
(or compared with)  a classical body  (\ref{classicalSG})  with  $N \sim O(10^{23})$. 

\subsection{An infinite spin puzzle}

Generally many spins inside a macroscopic body are oriented in random, different directions. But above all, the particles inside are bound in atomic, molecular and in crystaline structures. A bound particle does not split   \`a la Stern-Gerlach under an inhomogeneous magnetic field, because 
the bound-state Hamiltonian does not allow that  \footnote{A closely parallel observation is about the quantum diffusion.
Unlike free particles, particles in bound states (the electrons in atoms;  atoms in molecules, etc.) do not diffuse, as they move in binding potentials. This is one of the elements for the emergence of the classical mechanics, with unique trajectories  for macroscopic bodies. As for the center-of-mass (CM) wave function of an isolated macroscopic body, its free quantum diffusion is simply suppressed by  mass, see Table~\ref{diffusion}.}.

But what about a body made of many component spin $\frac{1}{2}$, {\it all} oriented in the same direction (e.g., a magnetized piece of metal)? 
Does such a body, with large spin, split in many sub wave packets in a strongly inhomogeneous magnetic field?        
The  question is  whether the three conditions recognized in \cite{KK2} for the emergence of classical mechanics  for a macroscopic body  with a unique trajectory,  reviewed in  Appendix~\ref{NewtonEq} here, are indeed sufficient.  Or,  is some extra condition, or a new unknown mechanism,  needed,  to suppress possible wide spreading  of the wave function into many sub-packets  
under an inhomogeneous  magnetic field?

\begin{figure}[t]
\begin{center}
\includegraphics[width=3.5in]{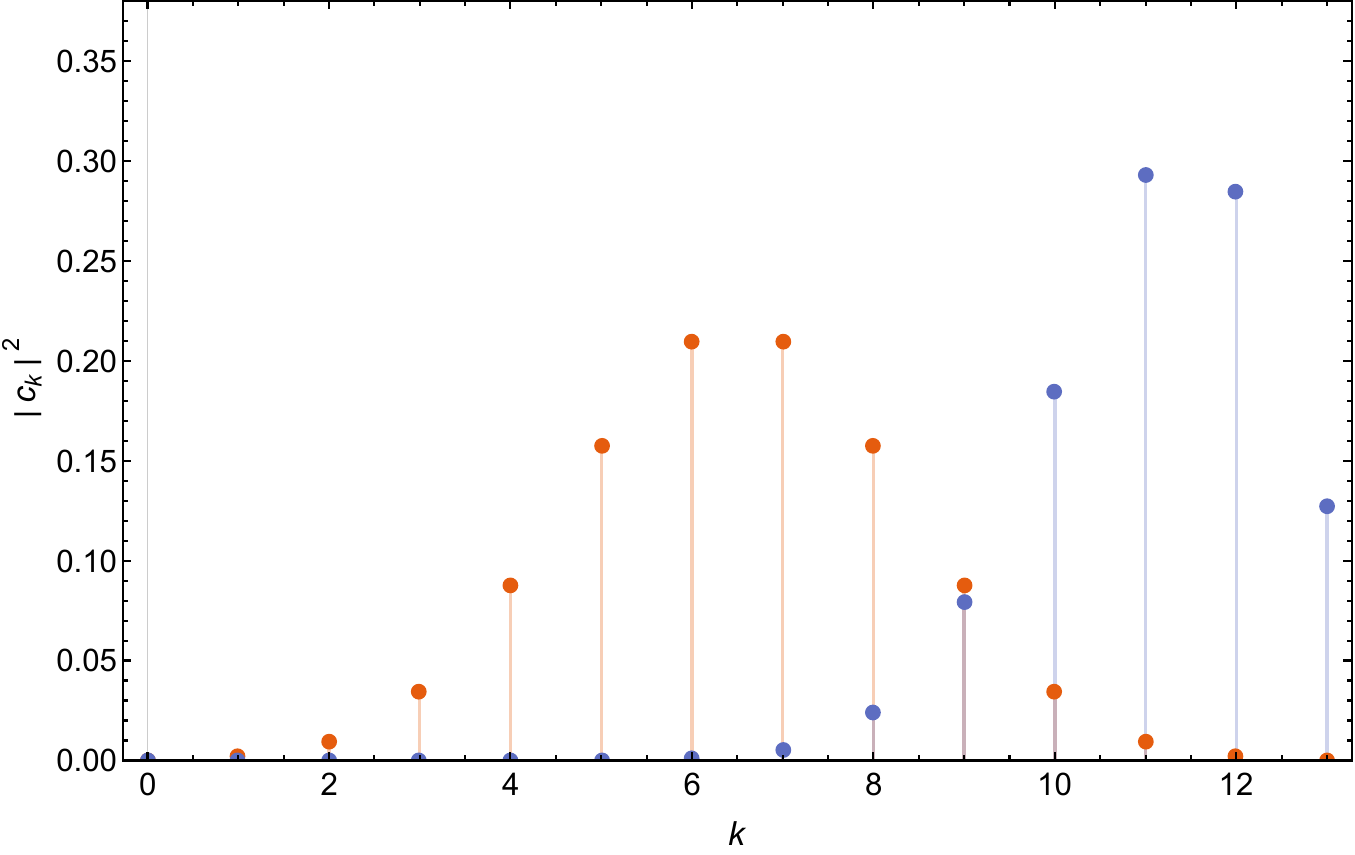}
\caption{\footnotesize {The distribution $|c_k|^2$ in $k$, i.e.~in possible values of $J_z=m$, $-j\le m \le j$  ($m = -j+k$)  for a spin $j=13/2$ particle in the state  (\ref{see00}), (\ref{exercise00}), with $\theta = \pi/2$ (center, orange dots) or with $\theta = \pi/4$  (right, blue dots).}}
\label{Spreadsi}
\end{center}
\end{figure}

The answer turns out to be simple, but somewhat unexpected \cite{KKHTE,KM}.   Consider the state of spin $j$, oriented towards a definite spatial direction, $  {\mathbf n} $, that is \footnote{These are known also as the Bloch state, or the spin coherent states in the literature \cite{Radcliffe, Puri, Arecchi, Aravind, Lieb}.}, 
 \be   {\hat  {\mathbf  J}}^2   \,  | j, {\mathbf  n}\ckt  =  j(j+1)   | j, {\mathbf  n}\ckt \,,\qquad  
 ({\hat  {\mathbf  J}}\cdot {\mathbf  n} ) \,  | j, {\mathbf  n}\ckt   = j\,  |j,  {\mathbf  n}\ckt \,,      \label{Bloch}  \ee   
where ${\mathbf n}$ is a unit vector directed towards  $(\theta, \phi)$ direction, 
\be    {\mathbf n} =     (\sin \theta \cos \phi,  \sin \theta \sin \phi, \cos \theta)\;.  
\ee
The projection of this state on various eigenstates of $J_z$  is give by
\be     
|j, {\mathbf  n} \ckt  = \sum_{k=0}^{2j}       c_k \,   |j, m \ckt\;,  \qquad  m = -j+k \;,    \label{see00} 
\ee
\be   
c_k =      {\binom{2j}{k}}^{1/2}   \,     e^{i (j-k) \phi } \left(\cos\tfrac{\theta}{2}\right)^{k}   \left(\sin\tfrac{\theta}{2} \right)^{2j -k} \;, \qquad   \sum_{k=0}^{2j}   |c_k|^2 =1\;, \label{exercise00}
\ee   
Under a magnetic  field with a strong gradient towards the ${\hat z}$ direction,   the wave function of a small spin particle will split in sub wavepackets, with relative weight proportional to  $|c_k|^2$,  ($j_z=  -j+k$),    as  in Fig.~\ref{WSG} (a)   for spin $\frac{1}{2}$, or  as in Fig.~\ref{Spreadsi} for a spin $\frac{13}{2}$ 
particle. 

\begin{figure}
\begin{center}
\includegraphics[width=4.0 in]{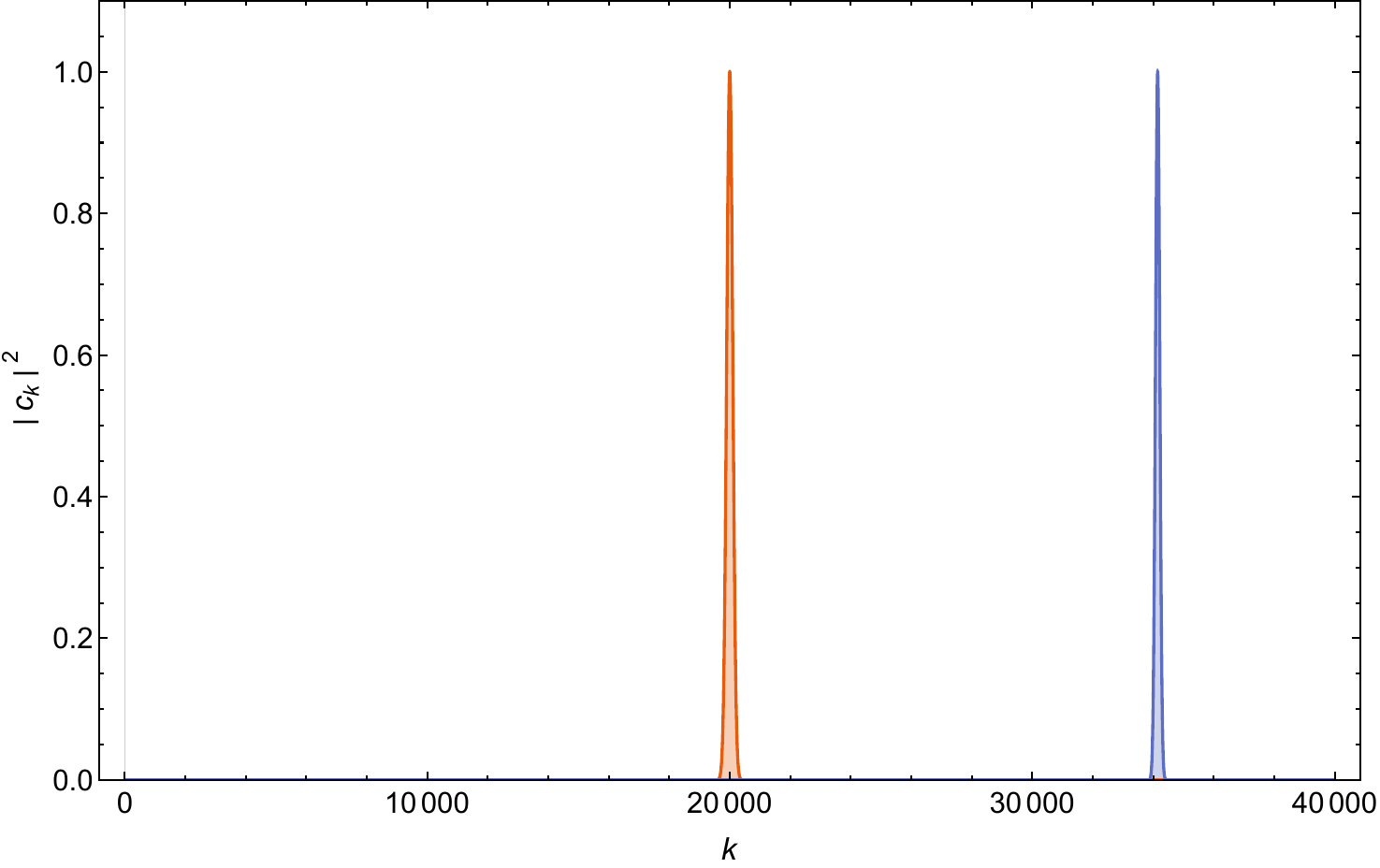}
\caption{\footnotesize The distribution $|c_k|^2$   ($J_z=m=-j+k$)   as in Fig.~\ref{Spreadsi} but for a spin $j  =2 \cdot 10^5$,  for $\theta = \pi/2$ (center, orange peak) or  $\theta = \pi/4$  (right, blue peak).  This figure is drawn by using the approximation (\ref{distr}) and  (\ref{distrLAM}), rather than the exact formula,  (\ref{see00}), (\ref{exercise00}). }
\label{Spreadno2}
\end{center}
\end{figure}

But the behavior for $j \to \infty$ turns out to be quite different.   See Fig.~\ref{Spreadno2},  for a spin $j=20000.$  Such a behavior can be understood by
using Stirling's formula in (\ref{exercise00}).   One finds, for  $n$ and $k=j+m$  both large  with $x = k/n$ fixed, the following  distribution in different values of $ m=-j+k$, 
 \be 
|c_k|^2   \simeq    e^{n  f(x)}\;, \qquad   x= k/n=(j+m)/2j \;,    \label{distr}    
\ee
where
\be  
 f(x)   =  - x \log x - (1-x) \log (1-x)    + 2 x  \log  \cos \tfrac{\theta}{2}  + 2 (1-x) \log   \sin \tfrac{\theta}{2}  \;.                \label{distrLAM}  
\ee
The saddle-point approximation  valid at $n \to \infty$, yields
\be   
f(x)    \simeq     - \frac{(x-x_0)^2 }{x_0(1-x_0)}   \;,  \qquad      x_0  =     \cos^2  \tfrac{\theta}{2}\;,  \label{spike1} 
\ee
and therefore 
\be    
\sum_k   |c_k|^2  (\cdots )     \longrightarrow      \int_0^1    dx   \,    \delta(x-x_0)  \, (\cdots )   \label{spike2} 
\ee
in the $n \to \infty$  ($x= k/n$ fixed)  limit. The narrow peak position $x=x_0$  corresponds to (see  Eq.(\ref{see00})) 
\be  
J_z =m  = n  (x-\tfrac{1}{2}) =   j  \, (2  \cos^2  \tfrac{\theta}{2}-1)  =   j \, \cos \theta\;.   \label{selection}
\ee
This means that a large spin ($j  \gg \hbar$) quantum particle with spin directed towards ${\mathbf n}$, in a Stern-Gerlach setting with an inhomogeneous magnetic field, moves along a single trajectory of a classical particle with  $J_z=   j  \, \cos \theta$,  instead of spreading over a wide range of split sub-packet trajectories covering  $-j  \le m   \le j$.  

This (perhaps) somewhat surprising result appears to indicate  that  quantum mechanics (QM)  takes care of itself,  so to speak,  in ensuring that a large spin particle   ($j/\hbar \to \infty$)  behaves classically, at least for these particular states $|j, {\mathbf n}\ckt $.     No extra conditions are necessary.   See \cite{KM}, however,  for more careful discussion on the quantum mechanical nature of generic large spin states, far from spin coherent states    $|j, {\mathbf n}\ckt$.

\subsection{Tunnelling molecules  \label{decoh2} }

\begin{figure}
\begin{center}
\includegraphics[width=5in]{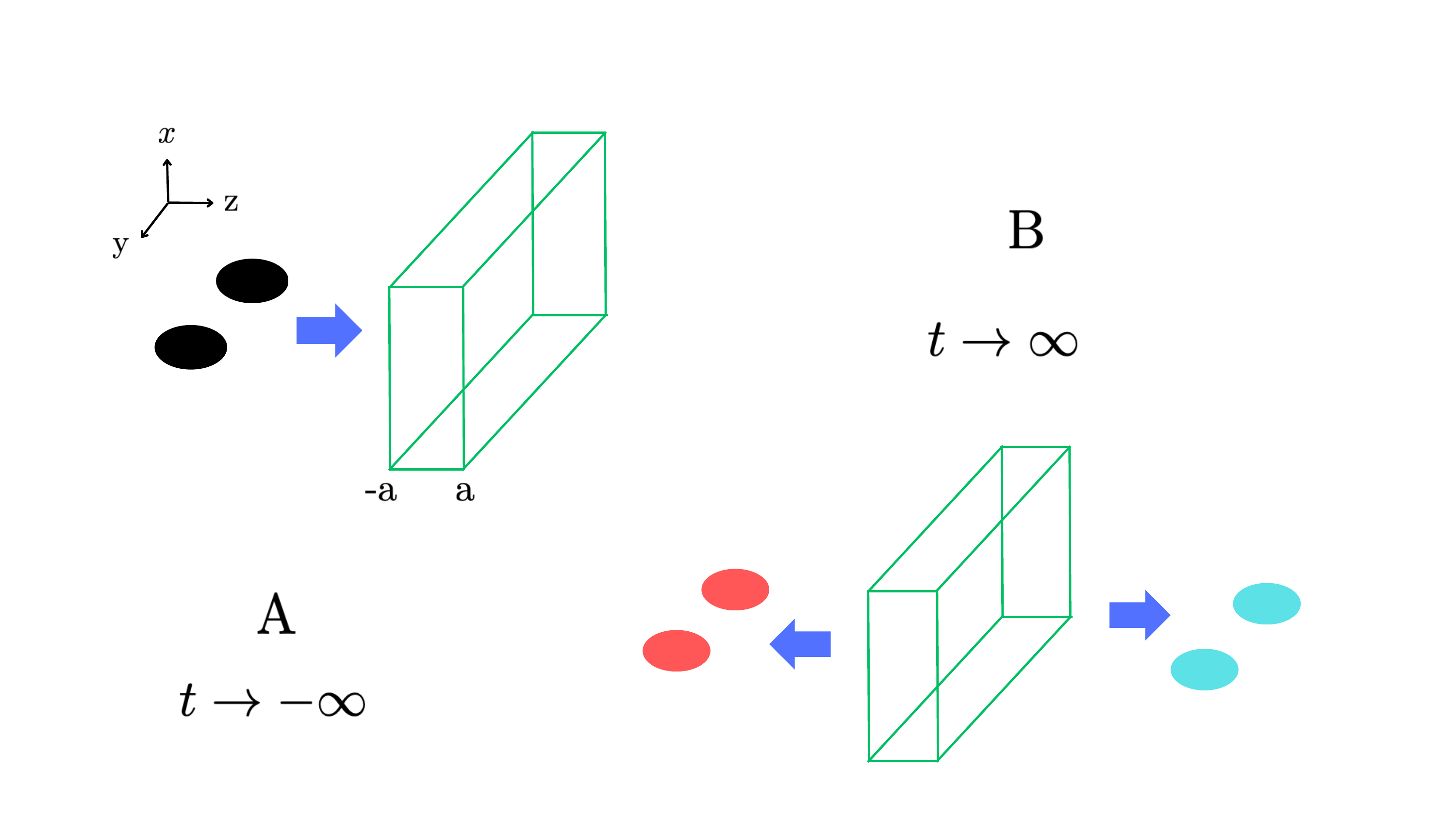}
\caption{\footnotesize  On the left figure (A),  an atom (molecule) arrives from $z=-\infty$ and moves towards the potential barrier   $V(z)$  at $ -a < z < a$  (independent of $x$ and $y$).   It is described by a wave packet (split in the transverse direction as in (\ref{Trwv})).   The wave function  of the particle   at $t \to \infty$,  shown in the right part (B),  contains both the reflected and transmitted waves. The coherent superposition of the two sub wavepackets in the $(xy)$ plane  remains intact.  
 }
\label{ET1}
\end{center}
\end{figure}

Another example of a process in which the distinction between decoherence and the classical limit can be neatly illustrated is 
a molecular (or atomic) beam, spilt in transverse direction $(x,y)$, 
 \be   \Psi   =    e^{i p_0  z / \hbar}   \psi(x, y)\;, \qquad     \psi(x,y) =   c_1  \psi_1(x,y) +    c_2  \psi_2(x,y)\;, \label{Trwv} 
 \ee
 where $\psi_1$ and $\psi_2$ are narrow (free) wave packets centered at   ${\bf r}_1=(x_1,y_1)$ and   ${\bf r}_2 =  (x_2,y_2)$, respectively.   
Actually,  we take a  wave packet, $\chi_{p_0}(p, z)$, also for the longitudinal wave function  by considering a linear superposition of the plane waves  $e^{i p   z / \hbar}$ with momentum $p$  narrowly distributed around  $p = p_0$.   For instance, a Gaussian distribution in  $p$,   $\sim e^{- (p-p_0)^2 / b^2}$,   will yield a Gaussian longitudinal wave packet in $z$ of width $ \sim 2 \hbar/b$. 
 At  times much less than the characteristic diffusion time    $t \ll   \frac{2m \hbar}{b^2}$,  the particle is approximately described by the wave function   \footnote{
 The exact answer has  the Gaussian width in the exponent replaced as 
 $\tfrac{b^2}{4\hbar^2} \to   \tfrac{b^2}{4\hbar^2  ( 1 +   i b^2 t/ 2 m \hbar)}$, and the overall wave function multiplied by  $ (1 +  i \, b^2  t  / 2m \hbar)^{-1/2}$.  
 These are  the standard  diffusion effects of a free Gaussian wave packet of width  $a =  2 \hbar/b$.  
 If the longitudinal wave packet and the transverse subwave packets are taken to be of a similar size,  then the free diffusion of the transverse wave packets 
 (hence $t$-dependence of $\psi(x,y)$) can also be neglected.
 },   
 \be     \Psi_{asymp}    \sim   e^{i p_0 z/\hbar}  e^{-  i p_0^2  t   / 2 m \hbar}  e^{- \tfrac{ b^2}{4\hbar^2} ( z -   \tfrac{p_0 t }{m})^2  }  \;    \psi(x, y)\;. 
 \ee
Assume that such a particle is incident from $z= -\infty$   ($t=-\infty$), moves towards right (increasing $z$),   and hits a potential barrier    (Fig.~\ref{ET1}),
 \be       V   =     \begin{cases}
   0 \;,   &     |z|  >    a , \\
    V(z)   \;,   &    - a   <  z  <   a 
\end{cases}
 \ee
 whose height is above the energy of the particle, approximately given by the longitudinal kinetic energy, $E \simeq   \tfrac {p_0^2}{ 2 m}$.    
 As the longitudinal and transverse motions are factorized,   the relative frequencies \footnote{It was proposed  in \cite{KK1,KKTalk}  to use ``(normalized)  relative frequency"  instead of the word  ``probability".  The traditional probabilistic Born rule places the human intervention at the center of its formulation, and distorts the
way quantum-mechanical laws   (the laws of Nature!) look. In the authors' opinion, this is  at the origin of innumerable puzzles, apparent contradictions and conundrums  
entertained in the past.  See  \cite{KK1,KKTalk}   for a new perspective and a more natural understanding of the QM laws. 
}    of finding the particle on both sides of the barrier
 (barrier penetration and reflection) at large $t$  can be calculated  by the standard one-dimensional QM. The answer is well known:  for instance the tunnelling frequency   is given, in the semi-classical approximation, by 
 \be   P_{tunnel} =  |c|^2\,,  \qquad  c  \sim   e^{ - \int_{-a_0}^{a_0}     dz  \, \sqrt{ 2m (V(z)-E)} / \hbar  }  \;,   \label{penetration}   
 \ee
  ($V(z)-E>0,\,\,   -a_0 < z < a_0$).  
   The particle 
  on the right  of the barrier  is described by the wave function
  \be    \Psi_{penetrated} \simeq    c\, \Psi_{asymp} =  c\,    e^{i p_0 z/\hbar}  e^{-  i p_0^2  t   / 2 m \hbar}  e^{- \tfrac{ b^2}{4\hbar^2} ( z -   \tfrac{p_0 t }{m})^2  }  \;    \psi(x, y)\;,  \label{right} 
  \ee
   where  $c$ is the transmission coefficient
   (\ref{penetration}).
  The transverse, coherent superposition of the two sub  wavepackets, (\ref{Trwv}),  remains intact.   See Fig.~\ref{ET1}.
  
\begin{figure}
\begin{center}
\includegraphics[width=5in]{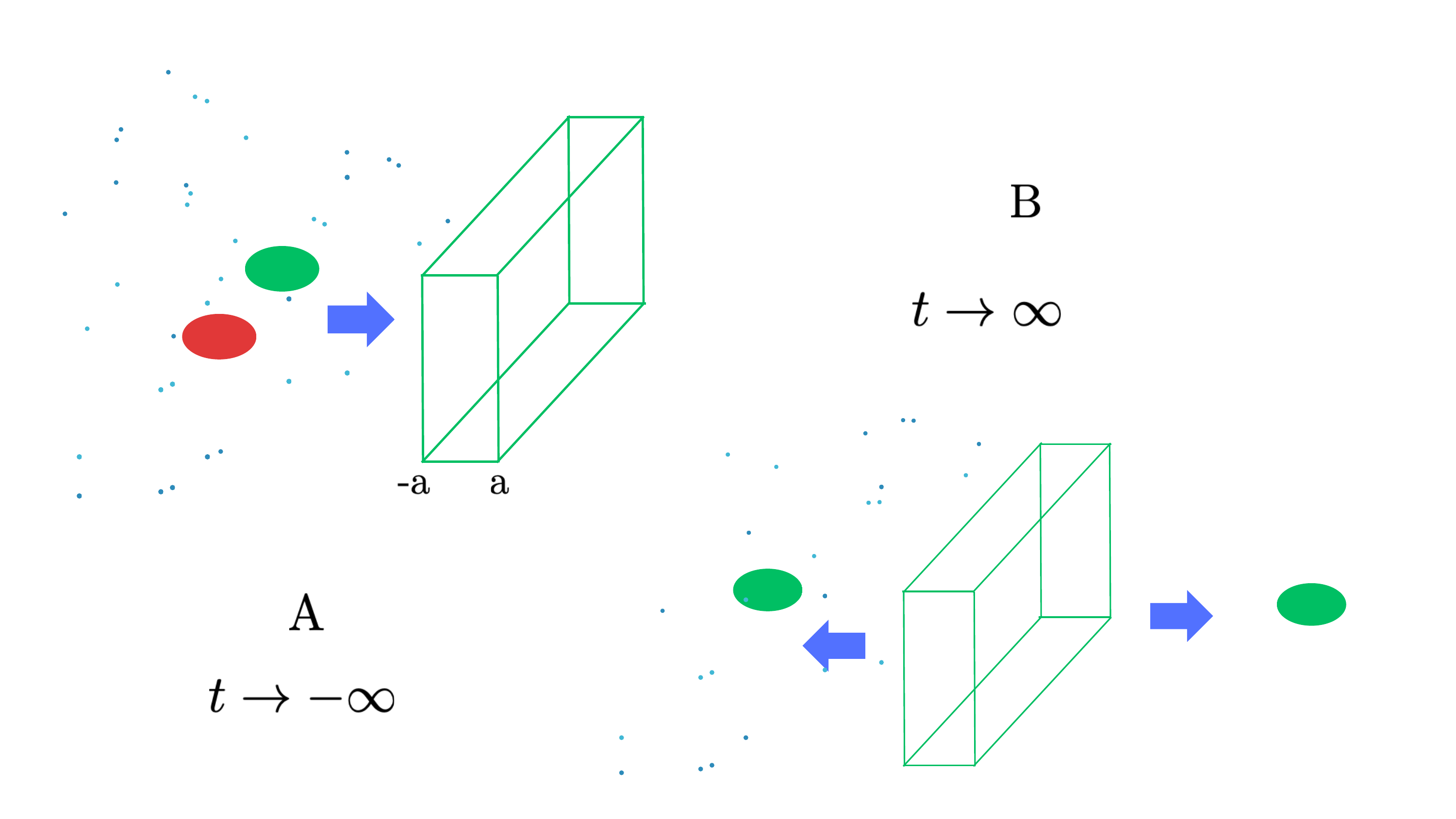}
\caption{\footnotesize  On the left figure (A), an atom (molecule) arrives from $z=-\infty$ and moves towards the potential barrier at $ -a < z < a$, as in  Fig.~\ref{ET1}.  But,  this time, the half space on the left of the  potential barrier contains air.  The molecule is now in a mixed state due to the environment-induced decoherence. Its (transverse) position density matrix became diagonal: it is either near  $(x_1,y_1)$  or   near  $(x_2,y_2)$.
  The wave function  of the particle   at $t \to \infty$,  shown in the right part (B), contains still a small transmitted wave as well as the reflected wave, however without  
coherent superposition of two transverse wave packets.  }
\label{ET2}
\end{center}
\end{figure}

Now reconsider the whole process, with the region left of the barrier  ($z < - a$) immersed in air.  The precise decoherence rate depends on several parameters,  but  the incident particles get decohered  in a very short time in general,  as in (\ref{mixed})  \cite{Joos1}-\!\cite{Zurek2}.   The particle at the left of the barrier \footnote{We assume that the environment particles  (air molecules)  have energy much less than the  barrier height, so that they are
confined in the region left of the barrier.}   is now a mixture:  each atom (molecule)  is either near  ${\bf r}_1=(x_1,y_1)$ or ${\bf r}_2 =  (x_2,y_2)$ in the transverse plane, just as in (\ref{mixed}). 
But when it hits the potential barrier it will tunnel through it, with the relative frequencies (\ref{penetration}), and will emerge on the other side of the barrier  as a free  particle.   It has the wave function, (\ref{right}),  with  $\psi(x,y)$ replaced by  $\psi_1(x,y)$,  with relative frequency  $|c_1|^2 / (|c_1|^2+ |c_2|^2)$, or  by  $\psi_2(x,y)$, with frequency    $|c_2|^2 / (|c_1|^2+ |c_2|^2$).  It is a statistical mixture, but each is a pure quantum mechanical particle. See Fig.~\ref{ET2}.

Our discussion here assumes that the air molecules are just energetic enough (their de Broglie wave length small enough) to resolve the transverse split wave packets   (see (\ref{mixed})), but are much less energetic than the longitudinal kinetic energy   $\tfrac {p_0^2}{ 2 m}$  and that  their flux is sufficiently small. In writing (\ref{right}) we assumed that the effects of the environment particles on the longitudinal wave packet  are small, even though the tunnel frequency may be somewhat   modified, as it is very sensitive to its energy.

Obviously, in a much warmer and denser environment the effects of the scatterings on our molecule would be more severe, and the tunnelling rate would become considerably smaller.  Even then, our atom (or molecule) remains quantum mechanical  \footnote{The situation is reminiscent of the $\alpha$ particle track  in a Wilson chamber.   $\alpha$ is scattered by atoms,  ionizing them  on the way,  but traces roughly a straight trajectory. 
When it arrives at the end of the chamber, it is just the same $\alpha$ particle.  It has not become a classical particle.}.

\subsection{Cosmic rays}

The cosmic rays (neutrinos, gamma, proton, etc.)  coming out of the  hot and dense
environments  of star's interiors,   once out,   propagate freely  in the intergalaxy  space  (a good approximation of the vacuum)  as pure-state quantum mechanical particles.

\section{Conclusion}

    The notion that the elementary particles are quantum mechanical, is usually  taken for granted in high-energy physics (and in general, physics) communities. However, as we are asking here  whether a molecule, a macromolecule, or larger particles, are quantum mechanical or classical,  and under which conditions,   it perhaps makes sense to ask whether or not the elementary particles are quantum mechanical, and if so, why.   
Introduction of the concept of the Quantum Ratio, and the related criterion,  allow us to answer at once  this question affirmatively, and  to explain why. 

    We are however not claiming that this is a new, original idea about the quantum mechanical nature of the elementary particles.  
Perhaps  one should go back to early $70'$ s when the standard model \cite{Weinberg,Salam,Glashow,GellMann} of the quarks and leptons
 has been established as the correct theory of the fundamental interactions.  The laws underlying the Nature seem to be written in terms of a
 unifying language of local, quantum field theory of nonAbelian gauge interactions \cite{tHooft}.  And these are relativistic, quantum  theories of 
 particles. 
 
 The fact that  elementary particles, thus electron and photon, and to certain extent the atoms and small molecules, are always quantum mechanical,  means 
  that even if the CM of a macroscopic body  behaves classically,  the internal microscopic degrees of freedoms continue to be quantum mechanical.  This is so, even if in a warm environment such as interiors of biological systems these particles will suffer from various sorts  of
 decoherence effects.   Decoherence however does not mean that the system affected becomes classical:  the latter becomes a mixture.   
 It is possible that certain quantum mechanical phenomena such as the tunnel effect survive decoherence, as discussed in  
 Sec.~\ref{decoh2}.  These questions constitute  one of the important research themes  in the nascent science of quantum biology  \cite{QuantumBiology}. 
 
 A bi-product of these considerations concerns  the abstract concept of ``a particle of mass $m$'',  familiar both in quantum-mechanics and  classical-mechanics textbooks, to formulate model systems such as a harmonic oscillator. The Quantum Ratio \cite{KKHTE}, and general ideas how  Newton's equations emerge from quantum mechanics for macroscopic bodies \cite{KK2},  tell us however  that  a model based on such an abstract concept of ``particle'',  without any information about its size $L_0$ and its composition, cannot be used to explain the emergence of classical mechanics. 
 

 In a recent attempt to  clean up our understanding of the so-called quantum measurement problems \cite{KK1},  a particular emphasis was given to  the particle nature of the fundamental entities of our world.  This is indeed the reason for the spacetime local (i.e., event-like) nature of any quantum measurement process at its core.  And this, combined with the unique classical state of matter (the reading) of the macroscopic measuring device after each measurement, explains what is often perceived as the  ``wave function collapse''.  
 
The Quantum Ratio  \cite{KKHTE} and the notion that the elementary particles are quantum mechanical, 
might have been thought as the final outcome of the series of considerations on the Great Twin Puzzles of Physics Today.
  It is heartwarming though  to realize that, actually, the idea of pointlike quantum nature of the fundamental entities of our world  was also at the very starting point \cite{KK1}  and characterizes the whole chain of reasonings  which followed \cite{KKTalk,KK3,KK2,KM}, and which has eventually led to the idea of the Quantum Ratio \cite{KK2,KKHTE}.

\section*{Acknowledgments}

The work by K.K.  is supported by the  INFN  special initiative grant, GAST (Gauge and String Theories).  K.K.   is especially grateful 
to Hans Thomas Elze for collaboration and for inviting him to participate and present this work at the 
11th International Workshop DICE2024, Castiglioncello (Tuscany).

\appendix

\section{Elementary particles \label{EP}} 

  The elementary particles  known today (as of the year 2024)  are the  quarks,  leptons (electron, muon, $\tau$ lepton),  the three types of neutrinos,  and the gauge bosons (the gluons, $W$, $Z$ bosons and the photon), plus the Higgs boson  (with  mass  $125,35 $ GeV/$c^2$),  with masses \cite{PDG},  
\begin{table}[h] 
  \centering 
 \begin{tabular}{|c|c|c|c|c|c|}
   \rule{0pt}{3ex}  $2.16 $  ($u$)   &   $4.67$ ($d$)     &   $93.4$ ($s$)    &  $1.27  \cdot 10^3$  ($c$)  &    $4.18 \cdot 10^3$  ($b$)  &    $172.7   \cdot 10^3$  ($t$)    \\
\end{tabular} 
  \caption{\footnotesize    The quark masses in  {MeV}/{$c^2$};     the errors not indicated.   $1$ MeV$/c^2   \simeq  
  1.782661 \cdot
  10^{-27}$ g    
   }\label{quarkmass}
\end{table}

\begin{table}[h] 
  \centering 
  \begin{tabular}{|c|c|c|c|}
   \rule{0pt}{3ex}   $0.51099895$  ($e$)  &   $105.658$  ($\mu$)    &   $1776.86$  ($\tau$)  &  $m_{\nu} \ne 0$\;;\quad   $m_{\nu} < 0.8$ eV$/c^2$         \\
\end{tabular} 
  \caption{\footnotesize The lepton masses. The $e$, $\mu$ and $\tau$  masses are given in MeV$/c^2$.       
   }\label{leptonmass}
\end{table}

\begin{table}[h]
  \centering 
  \begin{tabular}{|c|c|c|c|}
  \rule{0pt}{3ex}  photon  &  gluons   & ${W^{\pm}}$ (GeV$/c^2$)  &   ${Z}$ (GeV$/c^2$)    \\
  $0  $  &  $ 0$ &   $80.377\pm 0.012  $  &  $ 91.1876 \pm 0.0021 $       \\
  \end{tabular}
  \caption{\footnotesize Gauge bosons  and their masses  }
  \label{gbosons}
\end{table}

\section {Newton's equation for a macroscopic body  \label{NewtonEq}}  
The conditions needed for the CM of an isolated macroscopic body {\it  at finite body temperatures} to obey   
Newton's equations have been investigated in great care in   \cite{KK2}. They are 
\begin{itemize}
  \item[(i)] For macroscopic motions  (for which $\hbar \simeq 0$)  the Heisenberg relation does not limit the simultaneous determination -- the initial condition -- of the position  and momentum;
  \item[(ii)]   The absence of quantum diffusion, due to a large mass (a large number of atoms and molecules composing the body);  
  \item[(iii)]  A finite body temperature, implying the thermal decoherence and mixed-state nature of the body. 
\end{itemize}
Under these conditions, the CM of an isolated macroscopic body has a unique trajectory. 
Newton's equations for it  follow from the Ehrenfest theorem. See Ref.~\cite{KK2} for discussions on various subtleties and  for the explicit derivation of Newton's equation
under external gravitational forces,  under weak, static, smoothly varying external electromagnetic fields, and under a harmonic-oscillator potential.
Somewhat unexpectedly, the environment-induced decoherence \cite{Joos1}-\cite{Zurek2}  which is extremely effective in rendering macroscopic states in a finite-temperature environment  a mixture, is found not to be the most essential element for the derivation of classical mechanics  from quantum mechanics.

\section{A subtle face of the Stern-Gerlach experiment   \label{SGB}} 

 The Hamiltonian is given by
 \be   H=   \frac{{\bf p}^2}{2m}  + V\;, \qquad        V=  -  {\boldsymbol  {\mu}  \cdot {\bf B}}\;,\label{Hamiltonian}
\ee
\be    {\boldsymbol  \mu} =   \mu_B    \, g \,  {\mathbf s}\;,     \qquad \partial  B_z / \partial z \ne 0\;,  \label{Spin}
\ee
where $ \mu_{B} =  \frac{ e \hbar }{2 m_e c}$ is the Bohr magneton.  We recall the 
well-known fact that the gyromagnetic ratio $g \simeq 2$ of the electron and the spin magnitude $1/2$ approximately cancel,
so $\mu_B$  is the magnetic moment,
  in the case of the atoms such as $A_g$,  where a single outmost electron provides the total spin $\tfrac{1}{2}$. 

An example of the  inhomogeneous field ${\bf B}$ appropriate for the Stern-Gerlach experiment is  \cite{Platt,Alstrom} 
\be      {\bf B}  =   (  0,   B_y,  B_z ), \qquad   B_y = - b_0 \, y, \quad B_z = B_0 +  b_0  \, z\;     \label{magneticfield}
\ee   
which satisfy   $\nabla \cdot  {\bf B} =  \nabla \times {\bf B} =0$.  The constant field $B_0$ in the $z$ direction must be large, 
\be   |B_0|  \gg       |b_0 \, y|\;.  \label{largeB}
\ee
 in the relevant region of $(y,z)$ of the experiment.  
The wave function of the spin $\frac{1}{2}$ particle entering the SG magnet  has the form,   
\be    
  \Psi=  {\tilde \psi}_1({\bf r}, t) |\!\uparrow\ckt +   \, {\tilde \psi}_2({\bf r}, t)   |\!\downarrow\ckt \;.
\ee
obeying the Schr\"odinger equation,
\be   i \hbar  \frac{d}{dt}  \Psi =   H\, \Psi\;. 
\ee
By redefining the wave functions for  the upper and down  spin components  as
\be      {\tilde \psi}_1({\bf r}, t)  =   e^{i  \mu_B  B_0 t / \hbar}   \psi_1({\bf r}, t) \;, \qquad   {\tilde \psi}_2({\bf r}, t)  =   e^{-  i  \mu_B  B_0 t / \hbar}   \psi_2({\bf r}, t) \;,
\ee
  one finds that the up- and down- spin components  $\psi_{1}$ and  $\psi_{2}$  satisfy the separate  Schr\"odinger equations \cite{Platt} 
\be   i \hbar  \frac{\de}{\de t}   \psi_1 =     \left(    \frac{{\bf p}^2}{2m}   -      \mu_{B}  b_0    z    \right)       \psi_1 \;,\qquad  
  i \hbar  \frac{\de}{\de t}   \psi_2  =     \left(    \frac{{\bf p}^2}{2m}   +       \mu_{B}  b_0    z    \right)       \psi_2 \;.
     \label{SEq}
\ee
This is because 
the term   $\propto  -\mu_y B_y=- (g \mu_B  b)  s_y $  in the Hamiltonian (\ref{Hamiltonian})    
mixing the two components $\psi_{1,2}$  
  has acquired a rapidly oscillating phase factor, 
\be      \pm i  \mu_B\, b_0  y \,   e^{\mp   2  i  \mu_B  B_0 t / \hbar}   \;,
\ee
hence can be safely neglected.    
  The condition (\ref{largeB}) is crucial here.

  As explained in \cite{Alstrom},  this can be  classically understood as the spin precession  effect around the large constant magnetic field $B_0 \hat z$, thanks to which  the forces on the particle in the transverse ($\hat x, \hat y$) directions average out to zero  \footnote{
With a magnetic field $B_0$ of the order of $10^3$ Gauss, 
the precession frequency is of the order of $10^{11}$ in the case of the original SG experiment \cite{Alstrom}.  With the average velocity of $A_g$ atoms of the order of  $100 \, m/s$  and the size of the region of the magnetic field of about  a few $cm$ \cite{SG},  the 
timescale of the precession is orders of magnitude ($\sim 10^{-5}$)  shorter than the time the atoms spend in the region.}. The only significant force  it receives is due to the inhomogeneity in  $B_z$, (\ref{Spin}), which deflects the atom in the $\pm \hat{z}$ direction.

\end{document}